\documentclass[conference,letterpaper]{IEEEtran}
\usepackage{cite}
\ifCLASSINFOpdf
  \usepackage[pdftex]{graphicx}
  \DeclareGraphicsExtensions{.pdf,.jpeg,.png}
\else
  \usepackage[dvips]{graphicx}
  \DeclareGraphicsExtensions{.eps}
\fi
\usepackage{enumitem}
\usepackage{amsmath}
\usepackage{bm}
\usepackage{amsfonts}
\usepackage{amssymb}
\usepackage{bbm}
\usepackage{array}
\usepackage{algorithm}
\usepackage{algpseudocode}
\usepackage{pifont}
\usepackage{balance}

\newcommand\blfootnote[1]{%
  \begingroup
  \renewcommand\thefootnote{}\footnote{#1}%
  \addtocounter{footnote}{-1}%
  \endgroup
}

\def \c   {\mathbf{b}}  
\def \d   {\mathbf{c}}  

\def \hcl {\eta} 

\def \Angle   {\boldsymbol{\Omega}} 
\def \Anglea  {\theta} 
\def \Anglee  {\phi} 

\def \MapAng {\boldsymbol{\Psi}}
\def \MapAngx  {\alpha}
\def \MapAngy  {\beta}

\def \Maa {M}   
\def \Mae {M}   
\def \Naa {N}   
\def \Nae {N}   

\def \AR  {A}   
\def \wQ  {\mathbf{w}_{\Angle}}  
\def \wV  {\mathbf{w}_{\MapAng}} 
\begin{document}

\title{Beampattern-Based Tracking for Millimeter Wave Communication Systems }

\author{\IEEEauthorblockN{
Kang Gao\IEEEauthorrefmark{1},
Mingming Cai\IEEEauthorrefmark{1},
Ding Nie\IEEEauthorrefmark{1},
Bertrand Hochwald\IEEEauthorrefmark{1},
J. Nicholas Laneman\IEEEauthorrefmark{1},\\
Huang Huang\IEEEauthorrefmark{2} and
Kunpeng Liu\IEEEauthorrefmark{2}}
\IEEEauthorblockA{\IEEEauthorrefmark{1}Wireless Institute, University of Notre Dame\\
Email: \texttt{\{kgao, mcai,nding1, bhochwald, jnl\}@nd.edu}}
\IEEEauthorblockA{\IEEEauthorrefmark{2}Huawei Technologies Co., Ltd.\\
Email: \texttt{\{huanghuang, liukunpeng\}@huawei.com}}
}

\maketitle
\blfootnote{This work has been supported by Huawei Technologies Co., Ltd.}

\begin{abstract}
We present a tracking algorithm to maintain the communication link between a base station (BS) and a mobile station (MS) in a millimeter wave (mmWave) communication system, where antenna arrays are used for beamforming in both the BS and MS. Downlink transmission is considered, and the tracking is performed at the MS as it moves relative to the BS. Specifically, we consider the case that the MS rotates quickly due to hand movement. The algorithm estimates the angle of arrival (AoA) by using variations in the radiation pattern of the beam as a function of this angle. Numerical results show that the algorithm achieves accurate beam alignment when the MS rotates in a wide range of angular speeds.  For example, the algorithm can support angular speeds up to 800 degrees per second when tracking updates are available every 10 ms.
\end{abstract}



%
\IEEEpeerreviewmaketitle

\section{Introduction}

Millimeter wave (mmWave) bands have been proposed for enabling wideband high-speed wireless communication for next generation cellular systems.
Antennas operating in the mmWave bands are small. Therefore, a large number of antennas can be packed into arrays for portable devices, and the array can be used for beamforming to provide sufficient array gain to compensate for the high path loss of the mmWave channels \cite{bib:Marcus2005mmwave}.

Beamforming systems can be realized in several ways, such as analog implementation, digital implementation, and hybrid implementation of analog and digital \cite{bib:alkhateeb2014mimo}.
In mmWave systems, analog beamformer is more attractive because of its lower cost. Typically, analog phase shifters are employed to control a finite number of phase settings while preserving the magnitude.
We focus on beamforming using analog phase shifters, which is called analog beamforming.

The beams generated by analog beamformers are directional with narrow beamwidth, which is the angle range with array gain above a threshold\cite{bib:balanis2016antenna}.
In mmWave cellular systems, such directional beams can be used at both the base station (BS) and the mobile station (MS) to cover a certain range of angle of departure (AoD) or angle of arrival (AoA) with high array gain.
The establishment of the communication link is called initial access, and one example method of initial access is Beam Refinement Protocol (BRP) in the IEEE 802.11ad standard~\cite{bib:bg_7}.
In order to maintain the communication link, the directions of the beams have to be adjusted periodically according to the movement and rotation of the MS.
We are mainly interested in maintaining the link after its establishment.
However, without proper weight updates in the phase shifters, the link can be lost quickly.
According to \cite{bib:tsang2011detecting}, the angular speed of a MS can go up to $800^\circ$/s due to simple human hand movements.
In this case, if the weights are updated every 10ms, the beam at the MS can rotate $8^\circ$ between two updates. Thus the communication link could be seriously degraded or even disconnected, especially if beams of narrow beamwidth are used.

The weights should be updated periodically to align the beams with the varying AoD and AoA.
Efficient tracking algorithms are required, especially for the MS, because it can be rotated in a wide range of angular speed due to simple hand movements.
We focus on tracking at the MS in the downlink, using uniform planar arrays (UPA) at both the BS and the MS.
The method can also be generalized to the uplink, the BS, and antenna arrays with other geometry structure.

Several tracking algorithms have been proposed for analog beamforming.
In \cite{noh2015multi}, a codebook-based method is used for both link establishment and tracking.
A codebook is a set of codewords representing the weights for the phase shifters.
One codeword is selected from the codebook for each weight update, and the maximum achievable array gain is limited by the resolution of the codebook.
In \cite{bib:he2014millimeter}, tracking is based on a perturbation method.
The training beams are designed by perturbing the weights of the current beam, and the training beam providing the highest gain is selected for weight update.
However, this method only supports low MS angular speeds.

In this paper, we propose a beampattern-based tracking method, where
AoA is estimated based on the beampattern of the receiver array, determined by the geometry of the UPA.
The updated weights are created based on the estimated AoA.
Although our beampattern-based tracking algorithm requires more computations  than the perturbation method in \cite{bib:he2014millimeter}, it supports angular speeds up to $800^\circ$ per second.
Compared with the codebook-based method and the perturbation method, our beampattern-based tracking method has two advantages:
\begin{itemize}
\item It provides accurate beam alignment after each weight update, and achieves a near-optimal throughput.
\item It supports a wide range of angular speeds of MS rotation.

\end{itemize}


\section{system model} \label{System_Model}
\subsection{System Architecture}
\begin{figure}
\begin{center}
\includegraphics[width=3.3in]{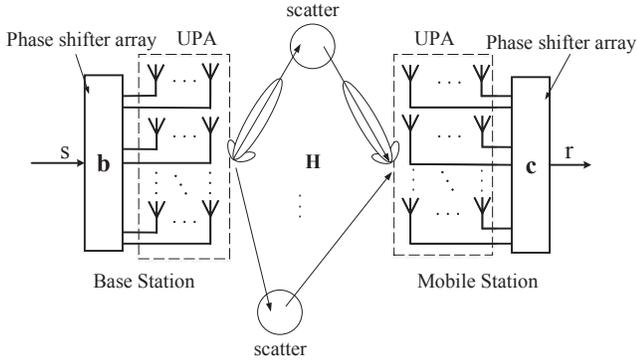}
\end{center}
\caption{A point-to-point downlink mmWave communication system with uniform planar arrays (UPAs). For simplicity, time and frequency indices of the signals are omitted.}
\label{fig:system_setup}
\end{figure}

We consider tracking in a point-to-point downlink wideband mmWave communication system with Orthogonal Frequency Division Multiplexing (OFDM). As shown in Fig. \ref{fig:system_setup}, both the BS and the MS have a single radio-frequency (RF) chain with a square UPA. Specifically, the BS has an $M\times M$ UPA while the MS has an $N\times N$ UPA. The distance between two adjacent antennas in both UPAs is $d$, and the wavelength of the carrier frequency is $\lambda$.
Our tracking algorithm can be generalized to the uplink, multiple RF chains, multiple users, non-square UPA, or narrow band communication.

The signal between the BS and the MS can be expressed as
\begin{equation}
r(f) = {\d}^H\mathbf{H}(f){\c} s(f) + {\d}^H\mathbf{z}(f),
\label{eq:channel_in_freq_domain}
\end{equation}
where $f\in\{0,1,\cdots, F-1\}$ denotes OFDM subcarrier indexes, $F$ is the total number of subcarriers, ${\c} = [b_1,\cdots,b_{M^2}]^T$ and ${\d}= [c_1,\cdots,c_{N^2}]^T$ are the beamforming weights of the BS and the MS, respectively, $s(f)$ and $r(f)$ are transmitted and received signals of subcarrier $f$, respectively, $\mathbf{H}(f)\in\mathbb{C}^{N^2\times M^2}$ denotes the frequency response of the channel matrix, and $\mathbf{z}(f)\in \mathbb{C}^{N^2\times 1}$ denotes the noise vector at the MS antenna array, whose entries are modeled as independent and identically distributed (i.i.d.) circularly-symmetric complex Gaussian random variables.

Because of hardware constraints, the magnitude of the beamforming weights are fixed and the angle is quantized into $Q$ bits. According to
\cite{bib:bakr2009impact}, the impact of quantization of the phases is small for phased array if $Q \geq 4$, which is true for a typical mmWave phase shifter. Therefore, we only focus on the magnitude constraint of magnitude in the design of weights,
i.e. $|{b}_i| = \frac{1}{M}$, $|{c}_j| = \frac{1}{N}$ for all $1\leq i \leq M^2$ and $1\leq j\leq N^2$.
We refer to such a constrained beamformer as a phase-shifter array.

\subsection{Channel model}
We focus on downlink tracking at the MS assuming that initial beam alignment has been established. According to \cite{bib:bg_8}, the non-line-of-sight (NLOS) path loss is much higher than that of the line-of-sight (LOS) path, implying the power decays significantly with reflections, and leads to a sparse channel with a few paths, which is also shown in \cite{bib:rappaport2013broadband}.
Furthermore, if directional beams of the transmitter and the receiver focus on one path, the other paths are filtered and the multipath effect is limited \cite{bib:yang2006frequency}. In our proposed tracking algorithm, we only consider the tracking of the dominant path in the channel model. Therefore, the channel is frequency-flat, and we can drop frequency $f$ in channel matrix for simplicity:
\begin{equation}
\mathbf{H} =  g\, \mathbf{a}_{R}(\Angle_{R})\,\mathbf{a}^H_{T}(\Angle_{T}),
\label{eq:cluster_channel_matrix}
\end{equation}
where $g$ is complex gain, $\Angle_{T} = [\Anglea_{T} ~ \Anglee_{T}]$ and $\Angle_{R} = [\Anglea_{R} ~ \Anglee_{R}]$ is the AoD and AoA of the tracked path, vectors $\mathbf{a}_{T}(\Angle_{T})\in\mathbb{C}^{M^2 \times 1}$ and $\mathbf{a}_{R}(\Angle_{R})\in\mathbb{C}^{N^2 \times 1}$ are the array responses of the BS and the MS at angle $\Angle_{T}$ and $\Angle_{R}$, respectively. Subscripts T and R denote transmitter and receiver, and $\Anglea$ and $\Anglee$ denote the azimuth angle and elevation angle, respectively.  According to \cite{bib:bg_3}, $\mathbf{a}_{T}(\Angle_T)$ and $\mathbf{a}_{R}(\Angle_R)$ can be expressed as

\begin{equation}
\begin{split}
\mathbf{a}_T(\Angle_T) = &\frac{1}{M} [  1, \cdots, e^{jp(m_T\cos(\Anglee_T)\sin(\Anglea_T) + n_T \sin(\Anglee_T))}, \cdots,\\
&e^{jp((\Maa-1)\cos(\Anglee_T)\sin(\Anglea_T) + (\Mae-1)\sin(\Anglee_T)) } ]^T,
\label{eq: BS_array_response}
\end{split}
\end{equation}

\begin{equation}
\begin{split}
\mathbf{a}_R(\Angle_R) = &\frac{1}{N} [  1, \cdots, e^{jp(m_R\cos(\Anglee_R)\sin(\Anglea_R) + n_R \sin(\Anglee_R))}, \cdots,\\
&e^{jp((\Naa-1)\cos(\Anglee_R)\sin(\Anglea_R) + (\Nae-1)\sin(\Anglee_R)) } ]^T,
\label{eq: MS_array_response}
\end{split}
\end{equation}
where $p=2\pi d\lambda^{-1}$, $\left(m_T, n_T\right)$ and $\left(m_R, n_R\right)$ are coordinates of an antenna element in the UPA coordinate system of the BS and the MS, respectively. $m$ denotes the horizontal index and $n$ denotes the vertical index. ${\d}^H\mathbf{a}_{R}(\Angle_{R})$ and $\mathbf{a}^H_{T}(\Angle_{T}){\c}$ show the array gain of the transmitter and the receiver, respectively.

Therefore, (\ref{eq:channel_in_freq_domain}) can be rewritten as
\begin{equation}
r(f) = \hcl({\c},{\d})s(f) + z_s(f),
\label{eq:channel_model_freq_domain}
\end{equation}
where
\begin{equation}
\hcl({\c},{\d}) = {\d}^H\mathbf{H}{\c},
\label{eq:equivalent_channel_between_chains}
\end{equation}
is the equivalent channel gain between the BS and MS RF chains, and is the product of the array gains of the transmitter and the receiver, $z_s(f)= {\d}^H\mathbf{z}(f)$ is the equivalent received noise. Since $\mathbf{z}(f)$ is formed of i.i.d complex Gaussian random variables, $z_s(f)$ is a complex Gaussian random variable.

For the received signal model in (\ref{eq:channel_model_freq_domain}), conventional algorithms like Maximum Likelyhood Estimator (MLE) \cite{bib:bg_4} can be used to estimate the channel $\hcl({\c},{\d})$ in an OFDM communication system.

\section{Beampattern-Based Tracking} \label{Tracking_Algorithm}
In our tracking algorithm, we perturb the beam of the MS to vary the channel gain for training.
The variation of the channel gain is proportional to the variation of the array gain or the radiation pattern, which can be used to estimate AoA.
The connection between the AoA and the gain at the corresponding angle in the radiation pattern will be discussed in this section.

\subsection{Tracking Setup} \label{Tracking_Setup}
\begin{figure}
\begin{center}
\includegraphics[width=3.3in]{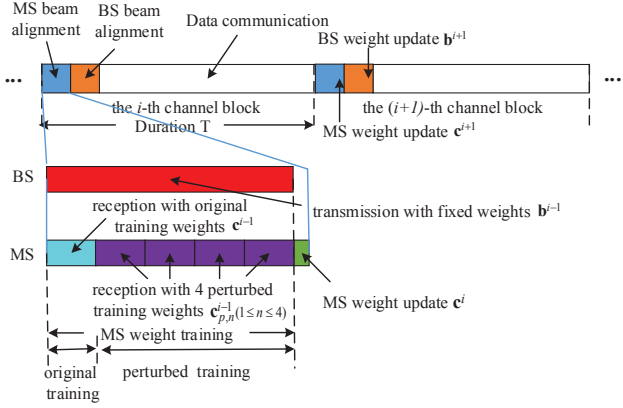}
\end{center}
\caption{Structure of channel blocks. Each channel block with duration T starts with MS beam alignment and BS beam alignment followed by data communication. At the end of MS/BS beam alignment, the weights of the MS/BS phase shifters are updated based on MS/BS weight training. During the MS weight training, the BS transmits signals with fixed weights of phase shifters, while the MS receives signals with five beams formed by five settings of phase shifters. BS beam alignment is similar to MS beam alignment.}
\label{fig:downlink_tracking_channel_block}
\end{figure}

The MS steers the beam to align the AoA by updating the weights of the phase shifters periodically with period $T$, the duration of a channel block. The structure of the channel blocks is shown in Fig. \ref{fig:downlink_tracking_channel_block}. It starts with MS beam alignment and BS beam alignment followed by data communication.
The weights of the MS/BS phase shifters are updated at the end of MS/BS beam alignment based on MS/BS weight training.
During MS weight training, the BS transmits signals with fixed weights received by the MS with five training beams using different training weights.
The first training weights are the weights obtained in last beam alignment, which are called original training weights.
The other four training weights are obtained by perturbing the original training weights,  and are called perturbed training weights.
The structure of BS beam alignment is similar.
Only 5 OFDM symbols are required for training and we assume that the channel does not vary much during the MS beam alignment.

We focus on tracking AoA at the MS and assume AoD does not change for simplicity. This happens when the location of the MS is fixed while the orientation of the MS antenna array changes due to the MS rotation. In this case, the weights of the BS are fixed without updating and denoted by $\c^0$.

Consider the MS beam alignment in the $i$th channel block.
Similar to $(\ref{eq:cluster_channel_matrix})$, the channel matrix of the tracked path during training is denoted as $\mathbf{H}^i$, with complex gain $g^i$, AoD $\Angle^i_{T}$ and AoA $\Angle^i_{R}$:
\begin{equation}
\mathbf{H}^i =  g^i\, \mathbf{a}_{R}(\Angle_{R}^i)\,\mathbf{a}^H_{T}(\Angle_{T}^i),
\label{eq:cluster_channel_matrix_time_block}
\end{equation}
The MS original training weights and MS perturbed training weights are $\d^{i-1}$ and $\d_{p,n}^{i-1}$ with beam focus angles $\Angle_{MS}^{i-1}$ and $\Angle_{p,n}^{i-1}$ ($1\leq n\leq 4$), respectively. The MS weights after update is denoted by $\d^i$, with beam focus angle $\Angle_{MS}^{i}$.

When the MS beam focuses on some direction $\Angle_{MS}$, the optimal weights of the MS phase shifters providing the highest gain is $ \d_{opt} = \mathbf{a}_R(\Angle_{MS})$, which is the array response at  $\Angle_{MS}$.
The physical explanation is that the weights of the phase shifters are designed to compensate the different phase delays of these elements when the signals come at a certain direction.
Therefore, the signals will be added constructively and reach a high gain. Mathematically, the maximum value of the channel gain $\hcl({\c},{\d})$ in (\ref{eq:equivalent_channel_between_chains}) is achieved when $\c = \mathbf{a}_T(\Angle_{T})$ and $\d = \mathbf{a}_R(\Angle_{R})$, where $\Angle_{T}$ and $\Angle_{R}$ are the beam focus angle at the transmitter and the receiver, respectively. When the phases of the phase shifters are quantized into $Q$ bits because of the hardware constraint, the weights vector becomes
\begin{equation}
\wQ(\Angle_{MS}): = \text{Quan}(\mathbf{a}_R(\Angle_{MS})),
\label{eq:MS_phase_shifter_weights}
\end{equation}
where $\text{Quan}\left(\cdot\right)$ is a quantization function which maps each element in a vector to the closest element in the quantization level to satisfy the hardware constraint. Equation (\ref{eq:MS_phase_shifter_weights}) shows the connection between the angle and the corresponding weights. The impact of quantization is small when $Q\geq 4$ \cite{bib:bakr2009impact}.

For the MS beam alignment in the $i$th channel block, $\Angle_{p,n}^{i-1}$ is generated based on $\Angle_{MS}^{i-1}$, which will be explained later, and $\d^{i-1}_{p,n}$ is calculated from $\Angle_{p,n}^{i-1}$ ($1\leq n \leq 4$) according to (\ref{eq:MS_phase_shifter_weights}). The corresponding channel gain ${\hcl}^i({\c^0},{\d})$ $(\d\in\{\d^{i-1},\d^{i-1}_{p,n}\})$ can be estimated using MLE \cite{bib:bg_4}, and is denoted as $\hat{\hcl}^i({\c^0},{\d})$. Based on the estimated channel gain, the AoA of the tracked path $\Angle_{R}^i$ can be estimated and assigned to
$\Angle_{MS}^i$, which will produce the updated weights $\d^i$.
For $i=1$, $\c^0$, $\d^0$ and $\Angle_{MS}^0$ are obtained in initialization for establishing the connection.

\subsection{Virtual Angle Mapping}
Virtual angle mapping is used in our tracking algorithm, and it helps to show the connection among the AoA, the MS beam focus angle, and the array gain. The mapping from an actual angle $\Angle = [\Anglea~\Anglee]$ to a virtual angle $\MapAng = [\MapAngx~\MapAngy]$ is defined as
$G_{MS}(\cdot): \Angle = [\Anglea~\Anglee] \rightarrow \MapAng = [\MapAngx~\MapAngy]$
\begin{equation}
\MapAngx = {\pi d \lambda^{-1}\Naa}\cos\Anglee\sin\Anglea, \quad
\MapAngy = {\pi d \lambda^{-1}\Nae}\sin\Anglee,
\label{eq: angle_mapping}
\end{equation}
where $\MapAngx$ and $\MapAngy$ are called virtual azimuth angle and virtual elevation angle, respectively.

Notice that there is a constraint on $\MapAng$,
\begin{equation}
{\MapAngx}^2 + {\MapAngy}^2 \leq (\pi d \lambda^{-1}N)^2.
\label{eq:angle_mapping_constraint}
\end{equation}

Inversely, the mapping $\Angle = G_{MS}^{-1}(\MapAng)$ is used to calculate the actual angle $\Angle$ from the virtual angle $\MapAng$,
\begin{align}
\Anglee = \arcsin\left(\frac{\MapAngy\lambda}{\pi d \Nae}\right);  \quad
\Anglea = \arcsin\left(\frac{\MapAngx\lambda}{\pi d \Naa\cos\Anglee}\right).
\end{align}

We can further get the connection between weights and virtual angle.
When the virtual angle of the beam focus angle $\Angle_{MS} = [\Anglea_{MS}~\Anglee_{MS}]$ is $\MapAng_{MS}= [\MapAngx_{MS}~\MapAngy_{MS}]$, the corresponding weights can be expressed as
\begin{equation}
\wV(\MapAng_{MS}): = \wQ(G_{MS}^{-1}(\MapAng_{MS})).
\label{eq:MS_phase_shifter_weights_virtual_angle}
\end{equation}

For a MS beam focusing on $\Angle_{MS}$, the magnitude of the array  gain at direction $\Angle_R = [\Anglea_{R}~\Anglee_{R}]$ with corresponding virtual angle $\MapAng_R = [\MapAngx_{R}~\MapAngy_{R}]$, can be expressed as
\begin{align}
\AR(\Angle_{MS},\Angle_R) &= \left| \wQ^H(\Angle_{MS}) \mathbf{a}_R(\Angle_{R}) \right| \label{eq:array_factor_def}\\
&\approx \left| \mathbf{a}^H_R(\Angle_{MS}) \mathbf{a}_R(\Angle_{R}) \right| \label{eq:quan_approximate}\\
&= \iota(\MapAngx_{MS}-\MapAngx_{R}, \Naa)\iota(\MapAngy_{MS}-\MapAngy_{R}, \Nae),
\label{eq:MS_array_factor}
\end{align}
where
\begin{align}
	\iota(\gamma, N)& = \frac{\sin(\gamma)}{N\sin(\frac{\gamma}{N})}  \label{eq:virtual_beam_width} \\
	& \approx \frac{\sin(\gamma)}{\gamma} \label{eq:virtual_estimate}
\end{align}
The approximation in (\ref{eq:quan_approximate}) follows from ignoring the quantization effect, and the approximation in (\ref{eq:virtual_estimate}) is achieved when  $\displaystyle \frac{\gamma}{N}\ll 1$.
The details of computation to obtain  (\ref{eq:MS_array_factor}) are omitted.

According to (\ref{eq:MS_array_factor}), the effect of the virtual azimuth angle and the virtual elevation angle on the radiation pattern are decoupled, and the magnitude of the gain only depends on the virtual azimuth/elevation angle offset. According to (\ref{eq:virtual_estimate}), the number of antennas in the array will not heavily influence the virtual beamwidth. The 3dB virtual beamwidth is about 2.8 for $N\geq 3$ through calculation of (\ref{eq:virtual_beam_width}).

When the AoA is fixed, the array gain only depends on the virtual angle offset, and the gain varies when the virtual beam focusing azimuth/elevation angle is perturbed.
The variation of the array gain, obtained through channel estimation, can be used to estimate the AoA.
Moreover, the virtual azimuth/elevation angle can be estimated independently, because their effects on the channel gain are decoupled in (\ref{eq:MS_array_factor}).
The steps of our algorithms are shown below.

\subsection{Beampattern-Based Tracking Method}

We denote the virtual angle of $\Angle_{MS}^{i-1}$, $\Angle_{p,n}^{i-1}$, and $\Angle_{R}^i$ by $\MapAng_{MS}^{i-1}$, $\MapAng_{p,n}^{i-1}$, and $\MapAng_{R}^i$.
The procedure of our beampattern-based tracking algorithm is shown as follows:


 \begin{enumerate}
\item Initialization: $i = 1$. Establish a line of communication, (for example, using BRP), which provides the initial setting of $\c^0$, $\d^0$, and  $\Angle_{MS}^0$.

\item Original Training: Set the weights of the BS and the MS phase shifters to be $\c^0$ and $\d^{i-1}$, respectively. The MS gets the estimation of the channel gain through MLE \cite{bib:bg_4}, denoted as $\hat{\hcl}^{i}(\c^0,\d^{i-1})$.

\item Design perturbed training weights $\d_{p,n}^{i-1}$ $(1\leq n \leq 4)$:
\begin{enumerate}
\item $\MapAng_{MS}^{i-1} = G_{MS}(\Angle_{MS}^{i-1})$ using (\ref{eq: angle_mapping}).

\item Obtain $\MapAng_{p,n}^{i-1}$ by perturbing $\MapAng_{MS}^{i-1}$
\begin{equation}
\MapAng^{i-1}_{p,n} = \MapAng_{MS}^{i-1} + \Delta\MapAng^{i-1}_{p,n},
\label{eq: training_angle_mapping}
\end{equation}
where
\begin{align}
\Delta\MapAng^{i-1}_{p,n} &= [\Delta\MapAngx^{i-1}_{p,n} \quad 0], (n\in \{1,2\}),\label{eq: def_delta_mapangle1}  \\
\Delta\MapAng^{i-1}_{p,n} &= [0 \quad \Delta\MapAngy^{i-1}_{p,n}], (n\in \{3,4\}),\label{eq: def_delta_mapangle3}
\end{align}
$\Delta\MapAngx^{i-1}_{p,n}$ and $\Delta\MapAngy^{i-1}_{p,n}$ are the perturbations of the virtual azimuth angles and elevation angles, respectively. Set $\Delta\MapAngx^{i-1}_{p,1} = \Delta\MapAngy^{i-1}_{p,3} = 0.7$ and $\Delta\MapAngx^{i-1}_{p,2} = \Delta\MapAngy^{i-1}_{p,4} = -0.7$. A smaller perturbation is selected if the constraint in (\ref{eq:angle_mapping_constraint}) is violated.

\item $\d_{p,n}^{i-1} = \wV(\MapAng^{i-1}_{p,n})$ using (\ref{eq:MS_phase_shifter_weights_virtual_angle}).
\end{enumerate}

\item Perturbed Training: Similar to original training in step 2), obtain $\hat{\hcl}^{i}(\c^0,\d_{p,n}^{i-1})$ for $1\leq n \leq 4$ with 4 trainings.

\item Estimate virtual angle offset between $\MapAng_{MS}^{i-1}$ and $\MapAng_{R}^i$:

\begin{enumerate}
\item Calculate $\hat{R}_{p,n}^{i-1}$ by
\begin{equation}
\hat{R}_{p,n}^{i-1} = \left|\frac{\hat{\eta}^i({\c}^0,{\d}_{p,n}^{i-1})}{\hat{\eta}^i({\c}^0,{\d}^{i-1})}\right|,
\label{eq: ratio_of_estimated_gain}
\end{equation}
\item Solve $\gamma \in (-\pi,\pi)$ in the following equations:
\begin{equation}
\hat{R}_{p,n}^{i-1} = \frac{\iota({\gamma}+\Delta\gamma^{i-1}_n,N)}{\iota({\gamma},N)}
\label{eq: estimated_gain_ratio}
\end{equation}
where
\begin{equation}
\Delta\gamma^{i-1}_n = \left\{
   \begin{aligned}
  &  \Delta\MapAngx^{i-1}_{p,n}, \quad n \in \{1,2\} \\   		
 &   \Delta\MapAngy^{i-1}_{p,n}, \quad n \in \{3,4\}
   \end{aligned}
  \right.
\label{eq: perturbation_step}
\end{equation}
and function $\iota({\gamma},N)$ is defined in (\ref{eq:virtual_beam_width}).

The solution of (\ref{eq: estimated_gain_ratio}) is denoted by $\hat{\gamma}^{i-1}_{n}$. Numerical methods can be used to solve the equation efficiently.

\item Obtain ${\Delta\hat{\MapAng}_{MS}^{i-1}}$, the estimate of the virtual angle offset
\begin{equation*}
{\Delta\hat{\MapAng}_{MS}^{i-1}} = [{\Delta\hat{\MapAngx}_{MS}^{i-1}} \quad {\Delta\hat{\MapAngy}_{MS}^{i-1}}],
\end{equation*}
where
\begin{align}
{\Delta\hat{\MapAngx}_{MS}^{i-1}} &= \hat{\gamma}^{i-1}_{n_1};  n_1 = \underset{1 \leq n \leq 2} {\mathrm{argmax}} ~ |\hat{\eta}^i({\c}^0,{\d}_{p,n}^{i-1})|, \label{eq: estimation_of_virtual_angle_x} \\
{\Delta\hat{\MapAngy}_{MS}^{i-1}} &= \hat{\gamma}^{i-1}_{n_2};  n_2 = \underset{3 \leq n \leq 4} {\mathrm{argmax}} ~ |\hat{\eta}^i({\c}^0,{\d}_{p,n}^{i-1})|.\label{eq: estimation_of_virtual_angle_y}
\end{align}
\end{enumerate}

\item Update $\Angle_{MS}^i$ and $\d^i$:
\begin{enumerate}
\item $\Angle_{MS}^i = G_{MS}^{-1}(\MapAng_{MS}^{i-1} - {\Delta\hat{\MapAng}_{MS}^{i-1}})$.

\item $\d^i = \wQ(\Angle_{MS}^i)$.
\end{enumerate}

\item Set $i=i+1$. Run step 2) for tracking in next MS beam alignment.
\end{enumerate}


\subsection{Explanation of the Algorithm} \label{Algorithm_Explain}
In the MS beam alignment of the $i$th channel block, $\hat{\hcl}^{i}(\c^0,\d^{i-1})$ and $\hat{\eta}^i({\c}^0,{\d}_{p,n}^{i-1})$ obtained in original and perturbed training are the estimates of ${\hcl}^{i}(\c^0,\d^{i-1})$ and ${\eta}^i({\c}^0,{\d}_{p,n}^{i-1})$, respectively.
Similar to (\ref{eq:cluster_channel_matrix}) and (\ref{eq:equivalent_channel_between_chains}), and according to (\ref{eq:array_factor_def}):
\begin{align}
	\left|{\hcl}^{i}(\c^0,\d^{i-1})\right| &= \left|\wQ^H(\Angle^{i-1}_{MS})\mathbf{H}^i{\c}^{0}\right| \nonumber\\
	&= \left|\wQ^H(\Angle^{i-1}_{MS})g^i\, \mathbf{a}_{R}(\Angle_{R}^i)\,\mathbf{a}^H_{T}(\Angle_{T}^i){\c}^{0}\right|  \nonumber \\
	&= \left| g_{BS}^i\right| A(\Angle^{i-1}_{MS},\Angle_{R}^i),\label{eq:cluster_gain_in_tracking}
\end{align}

where
\begin{equation*}
g_{BS}^i = g^i\mathbf{a}^H_{T}(\Angle_{T}^i){\c}^0.
\end{equation*}

Let $\MapAng_{R}^{i}=[\MapAngx_{R}^{i}~\MapAngy_{R}^{i}]$, $\MapAng_{MS}^{i-1}=[\MapAngx_{MS}^{i-1}~\MapAngy_{MS}^{i-1}]$, and $\MapAng_{p,n}^{i-1}=[\MapAngx_{p,n}^{i-1}~\MapAngy_{p,n}^{i-1}]$ be the virtual angles of $\Angle_{R}^i$, $\Angle_{MS}^{i-1}$, and $\Angle_{p,n}^{i-1}$, respectively. According to (\ref{eq:MS_array_factor}), equation (\ref{eq:cluster_gain_in_tracking}) becomes
\begin{equation}
\left|{\hcl}^{i}(\c^0,\d^{i-1})\right| = \left| g_{BS}^i\right|\iota(\MapAngx_{MS}^{i-1}-\MapAngx_{R}^i, \Naa)\iota(\MapAngy_{MS}^{i-1}-\MapAngy_{R}^i, \Nae)
\label{eq:channel_gain_of_one_cluster_original_training}
\end{equation}

Similarly, we have
\begin{equation}
\left|{\hcl}^{i}(\c^0,\d_{p,n}^{i-1})\right| = \left| g_{BS}^i\right|\iota(\MapAngx_{p,n}^{i-1}-\MapAngx_{R}^i, \Naa)\iota(\MapAngy_{p,n}^{i-1}-\MapAngy_{R}^i, \Nae)
\label{eq:channel_gain_of_one_cluster_perturbed_training}
\end{equation}

According to (\ref{eq: training_angle_mapping}), (\ref{eq: def_delta_mapangle1}), (\ref{eq: def_delta_mapangle3}), (\ref{eq:channel_gain_of_one_cluster_original_training}), and (\ref{eq:channel_gain_of_one_cluster_perturbed_training}) we have
\begin{equation}
{R}_{p,n}^{i-1} = \left|\frac{{\eta}^i({\c}^0,{\d}_{p,n}^{i-1})}{{\eta}^i({\c}^0,{\d}^{i-1})}\right| = \frac{\iota({\gamma}^{i-1}_{n}+\Delta\gamma^{i-1}_n,N)}{\iota({\gamma}^{i-1}_{n},N)}
\label{eq: ratio_of_actual_gain}
\end{equation}
where $\gamma^{i-1}_{n}$ is the virtual angle offset with $\gamma^{i-1}_{n}=\MapAngx_{MS}^{i-1}-\MapAngx_{R}^i$ for $n\in\{1,2\}$, $\gamma^{i-1}_{n}=\MapAngy_{MS}^{i-1}-\MapAngy_{R}^i$ for $n\in\{3,4\}$, $\Delta\gamma^{i-1}_n$ is the virtual angle perturbation defined in (\ref{eq: perturbation_step}). Also, (\ref{eq: ratio_of_actual_gain}) is a monotonic function of ${\gamma}^{i-1}_{n}$ for given $\Delta\gamma^{i-1}_n$, when ${\gamma}^{i-1}_{n}$ and $({\gamma}^{i-1}_{n} + \Delta\gamma^{i-1}_n) \in (-\pi,\pi)$.

$\hat{R}_{p,n}^{i-1}$ in (\ref{eq: ratio_of_estimated_gain}) is the estimate of ${R}_{p,n}^{i-1}$, and therefore the solution of equation (\ref{eq: estimated_gain_ratio}) $\hat{\gamma}^{i-1}_{n}$ is the estimate of ${\gamma}^{i-1}_{n}$. When the perturbation step $\Delta\gamma^{i-1}_n$ is small, the range of ${R}_{p,n}^{i-1}$ for ${\gamma}^{i-1}_{n}$ inside the beamwidth is compressed, which reduces the accuracy in the estimation of ${\gamma}^{i-1}_{n}$. When $\Delta\gamma^{i-1}_n$ is big, the perturbed training beam may be far away from the tracked path, decrease the array gain, and increase the influence of noise. $|\Delta\gamma^{i-1}_n| = 0.7$, which is $\frac{1}{4}$ of the virtual beamwidth is a suitable choice from simulations. The exact effect of the selection of the step requires further investigation.

As $\gamma_n^{i-1}$ for $n\in\{1,2\}$ and $n\in\{3,4\}$ estimate the offsets of the virtual azimuth and elevation angle, respectively, we select the ones with higher estimated channel gain, which implies higher SNR and less estimation error. The selection is shown in (\ref{eq: estimation_of_virtual_angle_x}) and (\ref{eq: estimation_of_virtual_angle_y}) in the algorithm.

\section{Numerical Result} \label{Numerical Result}
We consider a downlink OFDM communication between one BS and one MS with carrier frequency $73$ GHz (E-band available for ultra-high-speed data communication \cite{pi2011introduction}) and $2.5$ GHz bandwidth. One OFDM symbol has a period of $1\mu$s with $2048$ subcarriers and $341$ pilots.
Two UPAs with $M^2 = 16\times 16$ and $N^2=8\times 8$ antennas are used for the BS and the MS, respectively, with $d=\lambda/2$. There are $Q=4$ quantization bits for the BS and the MS phase shifters. The $3$ dB beamwidth of the MS UPA is about $13^\circ$ for both the azimuth angle and the elevation angle when the beam focuses on the broadside.

A channel model with two paths between the BS and the MS is used. One path is LOS and the other one is NLOS. The maximum SNR for the LOS and NLOS obtained in the MS are $5$ dB and $-8$ dB, when the beams of the BS and the MS are aligned with AoD and AoA of the LOS and NLOS, respectively. When the beam alignment for the LOS path is achieved, the NLOS path is filtered by the beam and has SNR$=-51.5$ dB. The location of the MS is fixed and it rotates along the azimuth angle with 3 different angular speeds $100^\circ/$s, $300^\circ/$s and $800^\circ/$s. The duration of each channel block $T=10$ms, the same as a frame duration in LTE \cite{bib:innovations2010lte}.

The available throughput $C(t)$ (bits/s/Hz) at time $t$ is used to show the performance.  $C(t)$ is expressed as
\begin{equation}
{C}(t) = \log_2\left(1 + \frac{({\d}(t))^H\mathbf{H}(t){\c}(t)}{\sigma^2}  \right)
\label{eq: throughput}
\end{equation}
where $\sigma^2$ is the noise power at the MS RF chain, ${\d}(t)$ and ${\c}(t)$ are the weights of the MS and the BS phase shifters, and $\mathbf{H}(t)$ is the channel matrix of the tracked path at time $t$.

Because the AoD is fixed in our scenario, $\c(t)$ is fixed as $\c^0$. Also, as the weights of the MS phase shifters are updated every $10$ms at the beginning of each channel block, we have ${\d}(t)=\d^{i(t)}$, where $i(t)=\lceil t/T \rceil$ indicating the indices of the channel blocks, and $\lceil \cdot \rceil$ is the ceiling function. Similar to (\ref{eq:cluster_channel_matrix}), $\mathbf{H}(t)=  g(t)\, \mathbf{a}_{R}(\Angle_{R}(t))\,\mathbf{a}^H_{T}(\Angle_{T}(t))$, where gain $g(t)$ and AoD $\Angle_{T}(t)$ are fixed, while AoA $\Angle_{R}(t)$ varies with time because of the rotation of the MS. In our example, $\Angle_{T}(0) = [0.1244 \quad -0.1235]$ and $\Angle_{R}(0) = [-0.7483 \quad 0.1235]$, with unit rads. For initialization, $\c^0$, $\d^0$, and $\Angle_{MS}^0$ are set based on beam alignment on LOS for $t=0$, i.e. $\c^0 = \wQ(\Angle_{T}(0))$, $\d^0 = \wQ(\Angle_{R}(0))$, $\Angle_{MS}^0 = \Angle_{R}(0)$.

Two other methods are used for comparison:

\textbf{Codebook method}:
According to \cite{noh2015multi}, we construct a codebook of UPA for MS with 128 codewords using Kronecker product of two different codebooks of uniform linear array (ULA), which are created using generalized DFT vectors.
The weights that provide the highest gain are selected from the codebook for the MS in each MS beam alignment.

\textbf{Perturbation method}:
According to \cite{bib:he2014millimeter}, we use nine training beams and select the training beam providing the highest gain as the updated beam in MS beam alignment in each channel block. The selection of the other parameters could be found in \cite{bib:he2014millimeter}.

\begin{figure}
\begin{center}
\includegraphics[width=3 in]{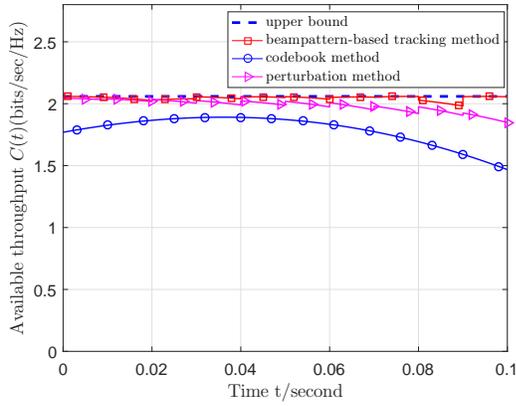}
\end{center}
\caption{Comparison of available throughput (\ref{eq: throughput}) for  $\omega=100^\circ$/s. The weights are updated every 10ms. Both the beampattern-based tracking method and the perturbation method provide high throughput. In the codebook method, the same weights are used after weight update unless a better codeword exists. In this figure, the same weights are used in the codebook method because of the limitation of the resolution, which provides a lower available throughput than the other methods.}
\label{fig:compare_100}
\end{figure}

\begin{figure}
\begin{center}
\includegraphics[width=3 in]{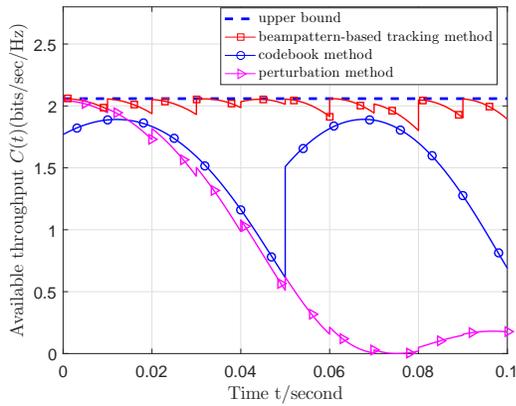}
\end{center}
\caption{Comparison of available throughput (\ref{eq: throughput}) for  $\omega=300^\circ$/s. The setup is the same as that in Fig.\ref{fig:compare_100} except that a different angular speed $\omega = 300^\circ$/s is used. Beampattern-based tracking method works well with near-optimal throughput after each weight update.  The perturbation method cannot follow the movement of AoA and the throughput drops quickly.}
\label{fig:compare_300}
\end{figure}

\begin{figure}
\begin{center}
\includegraphics[width=3 in]{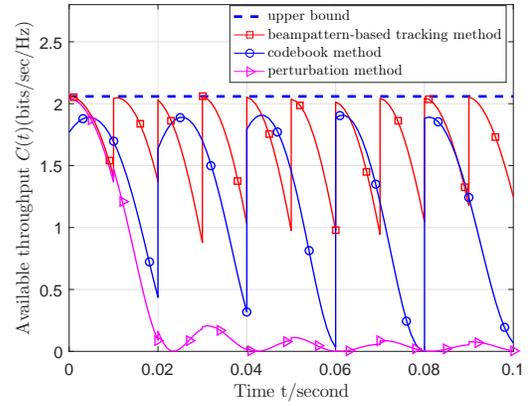}
\end{center}
\caption{Comparison of available throughput (\ref{eq: throughput}) for  $\omega=800^\circ$/s. The setup is the same as that in Fig.\ref{fig:compare_100} except that a different angular speed $\omega = 800^\circ$/s is used. Beampattern-based tracking method still works well with near-optimal throughput after each weight update. Both the codebook method and the perturbation method have lower available throughput.}
\label{fig:compare_800}
\end{figure}

The simulation results are shown in Fig. \ref{fig:compare_100},  \ref{fig:compare_300} and \ref{fig:compare_800}. The MS rotates with angular speed $\omega = 100^\circ$/s, $\omega = 300^\circ$/s, and $\omega = 800^\circ$/s, respectively.
We show the upper bound of throughput, when both the BS and the MS have perfect beam alignment all the same time, and the available throughput when these three methods (beampattern-based tracking method, the codebook method, and the perturbation method) are used for these 3 different $\omega$ during $0\leq t \leq 0.1$s.
From the three figures, we can see that the beampattern-based tracking method can provide the best beam alignment in tracking, and achieve near-optimal throughput after each weight update.
The throughput drops between weight updates because of the rotation of the MS. The codebook method has lower throughput than the beampattern-based tracking method because the accuracy of beam alignment is limited by the resolution of the codebook.
As a result, we can see that the codebook method does not update the weight effectively in Fig. \ref{fig:compare_100} and Fig. \ref{fig:compare_300}.
To reach a higher resolution, more codewords are needed, and more training is required to find the optimal codeword, which increases the the overhead.
The perturbation method works well for $\omega = 100^\circ$/s. However, the throughput drops quickly at high angular speeds for $\omega = 300^\circ$/s or $\omega = 800^\circ$/s, because it loses track.
The overhead of beampattern-based tracking method is small with only 5 training weights in each weight update, but 9 training weights are required in perturbation method.
In the codebook method, all of the codewords are tried during the training, but the performance is not as good as beampattern-based method.
One disadvantage of the beampattern-based tracking method is that it relies on the geometry of the antenna array, which requires calibration.
The effect of the mismatch between the actual array and the assumed array model needs further research.

\section{Conclusion} \label{Conclusion}
We proposed a beampattern-based tracking method for a mmWave communication system to track the MS and maintain the downlink between the BS and the MS. This method can be generalized to the tracking of the BS in downlink, and tracking of both the BS and the MS in uplink. No adaptive adjustment on the perturbation step is required in the algorithm. This method provides accurate beam alignment and achieves a near-optimal throughput after each weight update when the MS rotates in a wide range of angular speeds due to a simple hand movement. As our tracking method is based on the beampattern, determined by the geometry of the array, array calibration is required to provide good performance.



%
%
%



%

\bibliographystyle{IEEEtran}
\bibliography{IEEEabrv,IEEEexample,ref}
\balance

\end{document}